\documentclass[aps,prl,showpacs,amsmath,nofootinbib,
superscriptaddress,onecolumn,preprintnumbers]{revtex4-2}
\usepackage{natbib}
\usepackage{amsmath}
\usepackage{amssymb}
\usepackage{graphicx}
\usepackage{color}
\usepackage{slashed}
\usepackage[makeroom]{cancel}
\usepackage[normalem]{ulem}
\usepackage{slashed}
\usepackage{soul}
\usepackage{cancel}
\usepackage{paralist}
\usepackage{enumitem}

\begin{document}
\title{Extensive and Intensive Aspects of Astrophysical Systems and Fine-Tuning}
\author{Meir Shimon}
\affiliation{School of Physics and Astronomy, 
Tel Aviv University, Tel Aviv 69978, Israel}
\affiliation{Afeka Tel-Aviv Academic College of Engineering, Israel}
\email{meirs@tauex.tau.ac.il}


%

\begin{abstract}
Most astrophysical systems (except for very compact objects such as, e.g., black holes and neutron stars) in our Universe are characterized by shallow gravitational potentials, with dimensionless compactness $|\Phi| \equiv r_s / R \ll 1$, where $r_{s}$ and $R$ are their Schwarzschild radius and typical size, respectively. While the existence and characteristic scales of such virialized systems depend on gravity, we demonstrate that the value of $|\Phi|$ -- and thus the non-relativistic nature of most astrophysical objects -- arises from microphysical parameters, specifically the fine structure constant and the electron-to-proton mass ratio, and is fundamentally independent of the gravitational constant $G$.
In fact, the (generally extensive) gravitational potential becomes `locally' intensive at the system boundary; the compactness parameter corresponds to the binding energy (or degeneracy energy, in the case of quantum degeneracy pressure-supported systems) per proton, representing the amount of work that needs to be done in order to allow proton extraction from the system. More generally, extensive properties of gravitating systems depend on $G$, whereas intensive properties do not.
It then follows that peak rms values of large-scale astrophysical velocities and escape velocities associated with naturally formed astrophysical systems are determined by electromagnetic and atomic physics, not by gravitation, and that the compactness 
$|\Phi|$ is always set by microphysical scales -- even for the most compact objects, such as neutron stars, where $|\Phi|$ is determined by quantities like the pion-to-proton mass ratio. This observation, largely overlooked in the literature, explains why the Universe is not dominated by relativistic, compact objects and connects the relatively low entropy of the observable Universe to underlying basic microphysics. Our results emphasize the central but underappreciated role played by dimensionless microphysical constants in shaping the macroscopic gravitational landscape of the Universe. In particular, we clarify that this independence of the compactness $|\Phi|$ from $G$ applies specifically to entire, virialized or degeneracy pressure-supported systems, naturally formed astrophysical systems -- such as stars, galaxies, and planets -- that have reached equilibrium between self-gravity and microphysical processes. In contrast, arbitrary subsystems (e.g., a piece cut from a planet) do not exhibit this property; well within/outside the gravitating object the rms velocity is suppressed and $G$ reappears. Finally, we point out that a clear distinction between intensive and extensive astrophysical/cosmological properties could potentially shed a new light on the mass hierarchy and the cosmological constant problems; both may be related to the large complexity of our Universe.
\end{abstract}
\maketitle

\section{Introduction}\label{sec1}

The sensitivity of physical processes to the values of fundamental constants, such as the fine-structure constant $\alpha_{e}$ and the proton-to-electron mass ratio $\mu \equiv m_{p}/m_{e}$ -- where $m_{p}$ and $m_{e}$ are the proton and electron masses, respectively -- has been the subject of extensive theoretical and observational investigation. Variations in these constants impact not only particle physics, but also astrophysical processes, the formation of complex structures, and the potential habitability of the Universe.

Comprehensive reviews \cite{1,2} provide detailed accounts of how changes in $\alpha_{e}$ or $\mu$ modify atomic energy levels, nuclear reaction rates, and the epochs of recombination and primordial nucleosynthesis. For instance, even a modest increase in $\alpha_{e}$ accelerates atomic binding, shifting recombination to higher redshifts and potentially suppressing structure formation by altering the acoustic scale. Conversely, a lower $\alpha_{e}$ delays recombination and weakens atomic binding, with cascading effects on the cosmic microwave background (CMB) and baryon acoustic oscillations (BAO). Anthropic arguments have long explored the constraints imposed by these constants (e.g., \cite{3,4,5}). Other aspects have been investigated in, e.g., \cite{6,7,8,9,10}.

These and other explorations of how fundamental constants impact astrophysics and cosmology motivate us to question their possible -- if sometimes indirect or hidden -- role in shaping the velocity fields associated with typical astrophysical systems. Remarkably, velocities observed in such systems never exceed a threshold of a few thousand km/sec. For example, the Earth's velocity with respect to the Hubble frame is $\lesssim 400$ km/sec. The escape velocities from Venus, Earth, Mars, Saturn, and the Sun are approximately $6$, $11$, $5$, $35$, and $620$ km/sec, respectively. The solar system velocity relative to the Galactic center is $220$–$240$ km/sec, and typical infall velocities of merging clusters and subclusters have never been observed to exceed a few thousand km/sec. The corresponding gravitational wells are characterized by $\Phi + \beta^{2} \sim 0$, 
where $\Phi$ is the gravitational potential in speed of light-squared, $c^{2}$, units ($|\Phi|\equiv r_{s}/R$ is the ``compactness'' parameter where $r_{s}$ and $R$ are its Schwarzschild radius and actual size, respectively), and $\beta$ is the velocity in speed of light units.

This fact is somewhat puzzling, as a much more probable state would be a Universe teeming with 
black holes (BHs) and neutron stars (NSs), with $|\Phi| \lesssim 1$ and typical velocities approaching the speed of light. This is so simply because the next-to-highest entropy state of 
gravitating system is when it forms a BH, that eventually evaporates by emitting maximum entropy 
black body radiation.
Such a Universe would be thermodynamically closer to the final heat death state, in which all BHs are evaporated via Hawking radiation and the far-future asymptotic de Sitter dS space is filled with blackbody radiation characterized by the corresponding temperature $T_{dS} \sim 10^{-32}$ eV \cite{32}. Regardless of its initial state, the penultimate state of the Universe is expected to be populated by BH's and NS's. On small scales, such a Universe would be much less homogeneous than the one we actually observe; higher-entropy gravitational systems are simply more clustered. The puzzle, then, is why our Universe is so regular and homogeneous ($|\Phi| \ll 1$).
A key insight developed in this work is that, for entire, virialized (or supported by degeneracy pressure), and naturally-formed astrophysical systems -- such as stars, planets, and galaxies -- the dimensionless 
$|\Phi|$ is fundamentally set by microphysical parameters (such as $\alpha_{e}$, $\mu$, and other dimensionless ratios), and is independent of the value of the gravitational constant $G$; given the fact that $\alpha_{e}\ll 1$ and $\mu\gg 1$ the Universe is simply not sufficiently old for having all the potentially available nuclear fuel exhausted by now to allow for stars to collapse, the atomic scale that determine the mass density of rocky planets is much larger than nuclear scale, etc. In other words, non-gravitational interactions are sufficiently effective in inhibiting the Universe from reaching its ultimate fate even $\sim 14$ Gyrs after the big bang. The intensive nature of $|\Phi|$ does not apply to arbitrary subsystems (for example, a fragment cut from a planet), but holds for systems that have reached equilibrium through self-gravity and the relevant microphysical processes. Although the dependence of astrophysical masses and radii on fundamental constants is well explored, the general $G$-independence of $|\Phi|$ for virialized or degeneracy-pressured supported systems is not explicitly discussed in the literature, and we aim to highlight and clarify this point throughout the paper.

On a different front, and seemingly a sharp digression, it is worth noting that viewing the Planck mass $m_{pl}$ and gravitational masses $m_{gr}$ through the lens of extensive and intensive properties -- contrasted with the Higgs mass $m_{H}$ and inertial masses $m_{in}$ -- could potentially shed new light on the longstanding mass hierarchy problem, e.g. \cite{23,24,25,26}, 
spanning $\sim$17 orders of magnitude, as well as the cosmological constant problem, e.g \cite{27,28,29,30}, of $\sim$122 orders of magnitude. From this perspective, $m_{pl}^3$ and $m_{gr}^3$ may be regarded as extensive quantities, while the ratio $m_{gr}/m_{pl}$ remains intensive. Such a perspective leaves the standard gravitational framework intact while potentially offering fresh insights into why these fine-tuning 'problems' arise in the first place.

This paper is organized as follows. In Section \ref{sec2}, we summarize a few basic results from the literature on the typical masses, radii, and temperatures of certain virialized or degeneracy-pressure supported astrophysical systems.  
In Section \ref{sec3} we discuss the relation between extensive formulation 
of gravity and local scale invariance and how the latter emerges when intensive microscopic description is generalized to extensive formulation of gravity.
In Section \ref{sec4} we discuss possible implications of extensive Planck mass and 
we summarize in Section \ref{sec5}. Throughout, we adopt natural units where the Boltzmann constant $k_{B}$, Planck constant $\hbar$, and the speed of light $c$ are all set to unity. Our estimates are only crude order-of-magnitude calculations, but this is sufficient for our purposes.

\section{Dimensional Analysis Considerations}\label{sec2}

Dimensional analysis is a powerful technique that generally allows for order-of-magnitude estimates of key quantities, encapsulating the main physical ingredients in a transparent fashion without recourse to differential equations (e.g., \cite{11,12,13,14,15,16,17,18,19}). It captures the qualitative physics and underlying microphysics, leaving the details and precision to the full theory. 

In the following we start with the basic equations to correctly put the context and for future reference but quickly move on to dimensional considerations. The classical equations for gravitating systems include the Poisson equation for the gravitational potential $\Phi$,
\begin{eqnarray}
\nabla^2 \Phi = 4 \pi G\rho, 
\end{eqnarray}
and the continuity and Euler equations, respectively
\begin{eqnarray}
\frac{\partial \rho}{\partial t} + \nabla \cdot (\rho \mathbf{v}) = 0,\\
\frac{\partial \mathbf{v}}{\partial t} + (\mathbf{v} \cdot \nabla) \mathbf{v} = - \frac{\nabla P}{\rho} - \nabla \Phi,
\end{eqnarray}
supplemented by the equation of state (EOS)
\begin{eqnarray}
P=P(\rho, T, \{\chi_{i}\}). 
\end{eqnarray}
The latter relates the pressure, $P$, to the mass density, 
$\rho$, and possibly other parameters denoted by $\{\chi_i\}$. 
Here $\mathbf{v}$ is the velocity field.
In realistic systems $P$ is provided primarily by non-gravitational 
interactions or otherwise by quantum degeneracy. However, at the simple-minded 
level of discussion adopted here we model the astrophysical system as purely 
gravitational although the mass density values we adopt (as will be seen in the following) 
certainly rely on the existence of non-gravitational 
interactions. To support these over-idealized spherically-symmetric, constant density and static systems against their own gravitational pull without recourse to non-gravitationally 
produced pressure we are automatically lead to model our spherically 
symmetric static systems as de-Sitter solutions of the Einstein field 
equations in Section \ref{sec3.2}.

In the following, $E_{\star}$ is either the (electromagnetic) binding energy 
per atom, e.g. in the case of planets, nuclear burning stars, galaxies, etc., 
or the quantum degeneracy pressure in the case of WD's or NS's. 
The mass density and compactness of astrophysical system of 
mass $M$ and radius $R$, are given by
\begin{eqnarray}
M/R^{3} &\sim& \rho \\
GM/R &\sim& r_{s}/R \sim |\Phi|\sim E_{\star}/m_{p},
\end{eqnarray}
where the electromagnetic binding energy is consistent with the virial temperature of the system, $E_{\star}\sim T$, and $r_{s}\equiv 2GM$ is the Schwarzschild radius (strictly, it would be the corresponding de Sitter scale in our approximation but given that the current analysis is only an order of magnitude approximation the distinction is of no significant relevance). It should be stressed that whereas $\rho$ is a local property of the astrophysical system, the gravitational potential is not, and $|\Phi|\sim E_{\star}/m_{p}$ expresses values at the boundaries of entire, naturally-formed, systems, with typical size $R$. For such systems $E_{\star}/m_{p}\sim|\Phi|\sim G\rho R^{2}$ it follows that $R\propto G^{-\frac{1}{2}}$ provided that $\rho$ is indeed determined exclusively by microphysics. In this case $M\propto G^{-\frac{3}{2}}$, and so does $N\propto G^{-\frac{3}{2}}$ -- the number of protons which compose the system (assuming negligible binding energy). However, cutting a piece or fragment of size $\tilde{R}$ from the system, e.g. a rock from a planet, then whereas $\rho$ is unchanged 
$|\Phi(\tilde{R})|\sim G\rho \tilde{R}^{2}<G\rho R^{2}\sim|\Phi(R)|$ 
is clearly not G-independent for arbitrarily chosen $\tilde{R}$. It addition, the exterior 
solution for $r > R$ is $|\Phi|\sim GM/r=\Phi R/r$, and so the maximum value of the gravitational potential at $r=R$ is G-independent. The unique `status' of the system boundary that 
allows for the decoupling of $G$ from $|\Phi|$ at this surface is physically clear; typical 
velocities at the system boundary must not exceed the escape velocity. The latter is bounded by the thermal velocity which is in turn set by microphysics. 

In case that the compactness is determined by the binding energy, then since 
$|\Phi|=v_{esc}^{2}$, the requirement that the escape velocity $v_{esc}$ is larger than the thermal velocity $v_{th}$, and that 
the rms of the latter is determined by the temperature and the mass of the heaviest stable particle, the proton, via $m_{p}v_{th}^{2}\sim T$, we obtain that typically $|\Phi|\sim T/m_{p}$.
If both $\rho$ and $E_{\star}$ on the right-hand sides of Equations 5 and 6 are determined by the relevant microphysics of the system in question, then the system of Equations 5 and 6 can be solved for $R$, resulting in $R \sim \left(\frac{E_{\star} m_{p}}{\alpha_{g} \rho}\right)^{1/2}$ 
and $\alpha_{g} \equiv G m_{p}^{2}$. It then follows that $M \sim \rho \left(\frac{E_{\star} m_{p}}{\alpha_{g} \rho}\right)^{3/2}$. Crucially, Equation 6, $|\Phi|\sim E_{\star}/m_{p}$, implies that the 
compactness $|\Phi|$ -- not only the density $\rho$ -- is completely decoupled from $G$. The latter only affects the typical values of $M$ and $R$. As pointed out above, assuming $\rho$ 
and $E_{\star}$ are fully determined by microphysics and are independent of $G$, it immediately follows that $M \propto G^{-3/2}$ and $R \propto G^{-1/2}$. 
Exceptions are galaxies (and potentially galaxy clusters as well), whose typical $\rho$ does depend on $G$, and so $M$ and $R$ have different $G$-dependencies. Nevertheless, since even in these systems $T$ is independent of $G$, 
then $|\Phi|\sim T/m_{p}$ is independent of $G$, and the $G$-independence of the corresponding escape/virial velocities, maximum level of gravitational lensing and gravitational redshift automatically follows.

The temperature $T$ is a property of the system. For example, the temperature of planets cannot exceed the Rydberg energy (in fact, it must be much lower; see \cite{13}), the temperature of fuel burning stars cannot be smaller than $\sim 1$ keV, otherwise the Coulomb barrier cannot be overcome to allow for nuclear burning, etc. 
The idea is that temperature of virialized systems sets a lower 
limit on the escape velocity, and in case that the thermal velocity exceeds the escape velocity the system quickly disperses and disintegrates. However, the escape velocity sets the gravitational potential, i.e., the ratio $M/R$. If in addition $\rho \sim M/R^{3}$ is known from the microphysics of the system in question, then as argued above, these two constraints can be solved for the {\it characteristic} $M$ and $R$. It is perhaps worth mentioning that since $|\Phi|\sim E_{\star}/m_{p}$, then the highest possible temperatures of naturally formed astrophysical systems cannot exceed $10^{13}$K, otherwise $|\Phi|$ reaches unity, the escape velocity is close to the speed of light, and the system quickly disperses. For comparison, typical temperatures of the intracluster plasma in galaxy clusters fall in the range 
$\sim 10^{7}-10^{8}$K, accretion disks around BH's and NS's are usually not hotter 
than $10^{9}$K, the temperatures in cores of newly formed NS's are typically $10^{11}$K, etc.

Two immediate consequences of all this are as follows. First, 'line broadening' 
$\sigma_{\nu} = \sqrt{\frac{kT}{mc^{2}}}\nu_{0} \sim v \nu_{0}$ 
around a frequency $\nu_{0}$, caused by thermal motion of the emitters in a gravitationally 
bound system -- essentially a redshift smearing effect -- is 
maximized at the boundaries of astrophysical systems in a G-independent fashion. 
Second, although gravitational lensing is sourced by gravity, it is now trivial to see that 
the maximum deflection angle is independent of $G$. This simply follows from the fact that the geodesic equations depend only on $|\Phi|$, but the maximum absolute value of the latter is independent of $G$, and it then readily follows that the maximal deflection angle is independent of $G$. The latter could be a billion times larger or smaller than its measured value, 
yet the maximum value of the deflection angle would not change for it is set by 
$G$-independent microphysics. 
More formally, the deflection angle $\alpha$ is determined by the transverse 
impulse, i.e., the ratio of the transverse photon momentum change to its momentum, 
$\alpha = 2\int \nabla_{\perp}\Phi \, dl$. Now, since $\Phi$ is typically independent of $G$ in virialized and degeneracy-pressure supported astrophysical systems, then whatever $G$-dependence length scales might have, the combination $\alpha = 2\int \nabla_{\perp}\Phi dl$ remains independent of $G$ at its maximum.

In the following we inspect a few examples more closely. 
Our first example is rocky habitable planets. Following the arguments of \cite{13}, the temperature in habitable planets cannot exceed a certain threshold, specifically some fraction $\epsilon < 1$ of the Rydberg energy, $Ry \equiv \frac{m_{e}}{2} \alpha_{e}^{2}$, so as not to disrupt the chemistry required for sustaining living organisms, i.e., $|\Phi|\sim v_{esc}^{2} \lesssim T/m_{p}$ where $T = Ry$. All this implies, e.g., \cite{13}, that $|\Phi| \sim \epsilon \alpha_{e}^{2} \mu^{-1}$. In other words, since $T < 1$ eV and $m_{p} \lesssim 1$ GeV, then $|\Phi| = r_{s}/R \lesssim 10^{-9}$. For planet Earth, with radius $R \lesssim 10^{4}$ km, it implies that the corresponding Schwarzschild radius is $r_{s} \sim 1$ cm. The typical energy density of planets is given by $\rho \sim m_{p}/a_{0}^{3}$, where $a_{0} \equiv (m_{e} \alpha_{e})^{-1}$ is the Bohr radius, i.e., each Bohr volume is populated by approximately a single atom. Together, these simple atomic physics-based estimates of $T$ and $\rho$ uniquely determine $R$ and $M$ as described in Equations 5 and 6. In the case of planets in general, we require $T \sim Ry$ because the atomic structure of matter has to be preserved, albeit the chemistry of living organisms is no longer 
a condition that needs to be satisfied. Thus, we require $T \sim Ry$ and so $|\Phi| \sim \alpha_{e}^{2} \mu^{-1}$ in this case, and $R$ and $M$ are correspondingly larger than in the habitable planets case by factors of $\epsilon^{-1/2}$ and $\epsilon^{-3/2}$, respectively.

Another example is nuclear-burning stars. Based on very general nuclear kinematic arguments a typical stellar temperature $T \sim 0.1\alpha_{e}^{2}m_{p} \sim 10$ keV is deduced in \cite{14}. It then immediately follows that $|\Phi| \sim 0.1\alpha_{e}^{2} \sim 10^{-5}$. Here again, it is the weakness of the electromagnetic interaction that allows penetration through the Coulomb barrier and the ensuing nuclear fusion to take place. Had $\alpha_{e}$ been significantly larger, nuclear reactions would be impossible, resulting in matter collapsing under its own gravity into a black hole with $|\Phi| \sim 1$.

In \cite{15}, the estimate of galactic-scale systems required that for a gas cloud to fragment and form stars, the typical temperature has to drop below the Rydberg temperature. This by itself determines $|\Phi| \sim \alpha_{e}^{2} \mu^{-1}$. Another constraint used in \cite{15} was that the dynamical gravitational time is comparable to the (Thomson) cooling time. Together, these constraints resulted in a fair estimate of $M$ and $R$. However, this early analysis ignored the role of cold dark matter (CDM). It was revised later in \cite{16} to account for CDM. As expected, the conclusion that $|\Phi|$ is shallow is robust against this improved analysis.

Molecular clouds do not seem to have been discussed in the literature in the context of 
typical dimensional analysis-based scale estimates. Very general dimensional arguments imply that the vibrational and rotational molecular energy levels are given by $E_{vib} \sim \alpha_{e}^{2} m_{e} \mu^{-1/2}$ and $E_{rot} \sim \alpha_{e}^{2} m_{e} \mu^{-1}$. Considering virialized molecular clouds, the corresponding compactness parameters 
$|\Phi|$ are then $|\Phi| \sim \alpha_{e}^{2} \mu^{-3/2}$ and $|\Phi| \sim \alpha_{e}^{2} \mu^{-2}$, respectively, once again demonstrating the shallowness 
of $|\Phi|$. These estimates are general and decoupled from any assumptions pertaining to the corresponding $M$ and $R$. 

White dwarfs (WD) and neutron stars (NS) are held against their own gravity by degeneracy pressure. In the case of WD, the degeneracy is that of electrons with typical spacing of order the Compton wavelength of electrons, naively $T \sim m_{e}$, and the corresponding $|\Phi|\sim \mu^{-1}$. In the case of NS, the effective temperature is $m_{\pi}$, where $m_{\pi}$ is the pion mass, 
and so $|\Phi|\sim \kappa^{-1} \lesssim 1$, where $\kappa \equiv m_{p}/m_{\pi}$ 
is the proton-to-pion mass ratio. 
Thus, only when gravity is opposed by the strong interaction, or by no opposing force at all, 
is $|\Phi| \sim 1$ and the corresponding escape velocities become relativistic.

These results, along with others based on references \cite{13,14,15,16} for various 
virialized and degeneracy-pressure supported astrophysical systems, are compiled and summarized in Table \ref{table1}. The table presents typical masses, radii, number densities, and compactness in terms of the fine-structure constant $\alpha_{e}$, the proton-to-electron mass ratio $\eta \equiv m_{p}/m_{e}$, the proton-to-pion mass ratio $\kappa \equiv m_{p}/m_{\pi}$, the relative electromagnetic-to-gravitational interaction strength $\lambda \equiv \alpha_{e}/\alpha_{g}$ (where $\alpha_{g} \equiv G m_{p}^{2}$), and a dimensionless parameter $\epsilon \sim 0.1$. 

It is worth noting that, while the primary focus of the present work is the rightmost column showing the compactness parameter for a few typical astrophysical systems, the information in the other columns illustrates the valuable insights gained from basic dimensional analysis. For example, planets are less massive than their host stars fundamentally because $\alpha_{e} < 1$, NS's are less massive than their progenitor stars because $m_{\pi} < m_{p}$, and the observed Sun-to-Jupiter mass ratio, $O(10^{3})$, is simply determined by $\alpha_{e}^{-3/2}$. Likewise, the star-to-habitable planet (such as Earth) mass ratio is given by $\sim(\epsilon \alpha_{e})^{-3/2}$. While these observed ratios are typically explained by the details of various astrophysical processes, these processes are ultimately governed by dimensionless ratios of fundamental constants of nature, as demonstrated here. It is also important to realize that the fact that $|\Phi|\ll 1$ is not a result of the relative weakness of gravity in comparison to the other interactions gauged by, e.g. $\lambda\equiv\alpha_{e}/\alpha_{g}$, but rather a result of the weakness of the latter in absolute terms, i.e. the fact that $\alpha_{e}\ll 1$, $\mu\gg 1$, etc.

\begin{table}[h]
\begin{tabular} {|c |c|c|c|c|}
\hline
\hline
   & $M/m_{p}$ & $R/a_{0}$ & $n a_{0}^{3}$ & $|\Phi|$\\
\hline
\hline
 Habitable Planets & $\epsilon^{\frac{3}{2}}\lambda^{\frac{3}{2}}$ & 
 $\epsilon^{\frac{1}{2}}\lambda^{\frac{1}{2}}$ &  $1$ & $\sim\epsilon\alpha_{e}^{2}\mu^{-1}$\\
\hline
 Planets & $\lambda^{\frac{3}{2}}$ & 
 $\lambda^{\frac{1}{2}}$ &  $1$ & $\sim\alpha_{e}^{2}\mu^{-1}$\\
\hline
 Burning Stars & $\alpha_{e}^{-\frac{3}{2}}\lambda^{\frac{3}{2}}$ & 
 $10\alpha_{e}^{-\frac{3}{2}}\lambda^{\frac{1}{2}}\eta^{-1}$ & $(\alpha_{e}\mu/10)^{3}$&$\sim 0.1\alpha_{e}^{2}$\\
\hline
 Galaxies & $\lambda^{2}\alpha_{e}^{3}\mu^{\frac{1}{2}}$& $\alpha_{e}^{3}\mu^{\frac{1}{2}}\lambda$& 
 $\alpha_e^{-6}\lambda^{-1}\mu^{-1}$ & $\sim\alpha_{e}^{2}\mu^{-1}$\\
\hline
 White Dwarfs & $\alpha_{e}^{-\frac{3}{2}}\lambda^{\frac{3}{2}}$ & $\alpha_{e}^{\frac{1}{2}}\lambda^{\frac{1}{2}}$ &$\alpha_{e}^{-3}$& $\sim \mu^{-1}$\\
\hline
 Neutron Stars & $\alpha_{e}^{-\frac{3}{2}}\lambda^{\frac{3}{2}}[\kappa^{-3},\kappa]$& $\alpha_{e}^{\frac{1}{2}}\lambda^{\frac{1}{2}}\mu^{-1}$& 
 $(\mu/\alpha_{e})^{3}[\kappa^{-3},\kappa]$& 
 $[\kappa^{-3},\kappa$]\\
\hline 
\hline
\end{tabular}
\caption{A compilation based on \cite{13,14,15,16} of typical dimensionless values 
for various virialized and degeneracy-pressure supported astrophysical systems: 
typical masses in proton mass units $M/m_{p}$ (essentially $N$, the number of protons composing the system), radii in Bohr radius 
units $R/a_{0}$, number densities in Bohr units $na_{0}^{3}$ and the `compactness' 
parameter, i.e. the gravitational potential in $c^{2}$ units, $|\Phi|$. Throughout, 
$\mu\equiv m_{p}/m_{e}$ and $\kappa\equiv m_{p}/m_{\pi}$ are the proton-to-electron 
and propton-to-pion mass ratios, respectively, $\lambda\equiv\alpha_{e}/\alpha_{g}$ is the 
electromagnetic-to-gravitational coupling strength ratio, and $\epsilon\sim 0.1$.
In the case of NS a range of masses, densities and compactness are shown.
The table highlights that except for the case of galaxies $na_{0}^{3}$ (or $n$ for that matter) 
are independent of $G$. The compactness parameters, $|\Phi|$, are independent of $G$, and in fact $|\Phi|\lesssim\alpha_{e}^{2}$ for various systems spanning a range of mass and length scales. \label{table1}} 
\end{table}

As discussed above and further demonstrated in Table \ref{table1}, gravity, as encapsulated by the gravitational constant $G$, primarily determines the overall masses, radii, and particle numbers of gravitationally bound systems, setting the scale for the existence of macroscopic astrophysical objects. However, once such systems form and reach virial equilibrium, the maximum depth of their gravitational wells, as characterized by the dimensionless compactness $|\Phi|$, is dictated by microphysical parameters and remains fundamentally independent of the value of $G$. Whereas the peak values of the rms velocities in and around these systems 
is $G$-independent these velocities do depend on $G$ well within or outside the system boundary.
In this sense, $G$ enables the presence of large-scale structures, but does not directly govern 
the maximum characteristic velocities or compactness of these virialized systems.

In hindsight, the $G$-independence of $|\Phi|$ becomes even more transparent when derived directly from Equation 5, since $G$ does not enter the derivation; both $E_{\star}$ and $m_{p}$ are evidently independent of $G$. The results of this direct approach can be contrasted with the estimates in the rightmost column of Table \ref{table1} for comparison and verification. For instance, the typical binding energy of rocky planets is the Rydberg energy, 
$Ry \sim \alpha_{e}^{2} m_{e}$, as they are composed of atomic matter. 
Consequently, $|\Phi| \sim Ry / m_{p} \sim \alpha_{e}^{2} \mu^{-1}$ in this case. 
Compactness parameters $|\Phi|$ are shown for reference in Table \ref{table2} for planets in our 
solar system. Similarly, gas cloud fragmentation in star-forming galaxies depends on the gas temperature, with the Rydberg energy serving as a threshold \cite{15}, implying that typical galactic compactness is comparable to that of planets, $|\Phi| \sim \alpha_{e}^{2} \mu^{-1}$. For habitable planets, the binding energy is further reduced by a factor $\epsilon$, as discussed in \cite{13}. In the case of WD's, the relativistic electron degeneracy pressure balancing gravity corresponds to an energy scale $E_{\star} \sim p_{\star} \sim 1/d \sim m_{e}$, 
where $p_{\star}$ is the typical electron momentum uncertainty consistent with Heisenberg’s principle, and $d$ is the electron Compton wavelength, $m_{e}^{-1}$. Thus, $|\Phi| \sim \mu^{-1}$. For NS's, the non-relativistic degeneracy pressure yields $E_{\star} \sim p_{\star}^{2}/m_{p}$, 
with $p_{\star}\sim m_{\pi}^{-1}$, resulting in $|\Phi| \sim \kappa^{-2}$, which lies within the range $[\kappa^{-3}, \kappa]$ shown in Table \ref{table1}.

Since $|\Phi|$ at the system boundary can be inferred from fundamental physics 
directly, with no reference to estimates of $M$ and $R$, i.e. with no reference to $G$, 
the possibility of reformulating gravitation in a $G$-independent fashion -- 
for that purpose only -- could be useful. Consider the classical gravitational attraction between 
two masses $m_{1}$ and $m_{2}$. The mutual attraction force $F=Gm_{1}m_{2}/r^{2}$ 
can be cast in the form $F=m_{1}m_{2}/r'^{2}$ where $r'\equiv r/l_{pl}$ is the 
relative distance in Planck units, and $l_{pl}\equiv\sqrt{G}$ is the Planck length. 
This $G$-independent description, where 
distances are measured in Planck units rather than standard meters, inches, etc., is especially 
suitable for a $G$-free description of $|\Phi|$. We demonstrate in Section \ref{sec3.3} below 
how such a reformulation of GR can be achieved.

\begin{table}[h]
\begin{tabular} {|c |c|c|c|}
\hline
Planet & M (kg) & R (km) & $|\Phi|$ \\
\hline
Mercury & $3.30 \times 10^{23}$ & 2439.7 & $1.00 \times 10^{-10}$ \\
Venus   & $4.87 \times 10^{24}$ & 6051.8 & $5.98 \times 10^{-10}$ \\
Earth   & $5.97 \times 10^{24}$ & 6371.0 & $6.96 \times 10^{-10}$ \\
Mars    & $6.42 \times 10^{23}$ & 3389.5 & $1.41 \times 10^{-10}$ \\
Jupiter & $1.90 \times 10^{27}$ & 69911.0 & $2.02 \times 10^{-8}$ \\
Saturn  & $5.68 \times 10^{26}$ & 58232.0 & $7.24 \times 10^{-9}$ \\
Uranus  & $8.68 \times 10^{25}$ & 25362.0 & $2.54 \times 10^{-9}$ \\
Neptune & $1.02 \times 10^{26}$ & 24622.0 & $3.08 \times 10^{-9}$ \\
\hline
\end{tabular}
\caption{M, R and $|\Phi|$ of planets in our solar system. 
The compactness parameter $|\Phi|$ is always $O(10^{-9})$ or smaller 
in case of rocky planets (which matches the expectation that 
$|\Phi|\sim \alpha_{e}^{2}\mu^{-1}$ and 
$|\Phi|\sim\epsilon\alpha_{e}^{2}\mu^{-1}$ for planets and habitable 
planets, respectively, and is higher than (or about) that threshold 
value in case of the gaseous planets: Jupiter, Saturn, Uranus and Neptune. 
The latter simply do not have the strength of solids to resist compression under their 
own gravity. Consequently, their radii $R$ are not as large as those of rocky planets 
compared to their respective Schwarzschild radii, thereby resulting in correspondingly 
larger $|\Phi|$ values.\label{table2}}
\end{table}

\section{Microphysically-Determined Self-Gravitating Systems and the Independence from 
$G$}\label{sec3}

While this section establishes the general mathematical framework demonstrating the independence of $\rho$, $\Phi$ and $v$ from the gravitational constant $G$, it is in the 
context of virialized astrophysical systems and degeneracy supported-pressure that these 
insights find their concrete physical realization. 
We build the formalism in three stages until we arrive at 
a framework that relates the macroscopic description that allows for the existence of extensive quantities to an all-intensive microscopic-frame description via WI breaking and coordinate relabelling that results in dimensionless scalar fields and dimensionless length units where 
$G$ is nowhere to be found.

\subsection{G-free Description of Naturally-Formed Isolated Astrophysical Systems}\label{sec3.1} 

Consider a class of self-gravitating systems in which both the mass density, $\rho$, and the compactness parameter at the boundary of naturally-formed astrophysical systems, $|\Phi|$, are determined entirely by microphysical parameters -- specifically, by atomic and particle constants such as the proton mass $m_p$. For such systems, the typical compactness parameter can be written as $|\Phi|\sim\frac{E_{\star}}{m_p}$, as in Equation 5, where $E_{\star}$ is either the binding energy or degeneracy energy per proton. 

To remove the dependence of observable phenomena on the gravitational constant $G$, we relabel spatial and temporal coordinates defined by ${\bf x}\rightarrow {\bf x}/\sqrt{G}$ and 
$t' \rightarrow t/\sqrt{G}$. 
In these primed coordinates the structure of Eqs. (2)-(4) remains unchanged (with all differentiations are with respect to the primed coordinates), and Equation 1 is modified to
\begin{eqnarray}
\nabla'^{2}\Phi=4\pi G^{2}\rho(G)=4\pi\rho(G)/\rho_{pl}, 
\end{eqnarray}
where $\rho$ is now allowed to depend on $G$, a step which will be motivated 
below, and $\rho_{pl}=m_{pl}^{4}$ is the Planck energy density -- Planck mass per Planck volume. 
The operator $\nabla'$ denotes differentiation with respect to the rescaled dimensionless primed coordinates, and all quantities are expressed in terms of the latter, i.e. microphysical units.

As discussed above, we assume that $\rho$ is determined by microphysics. For example, 
$\rho\sim m_p/(2a_0)^3\propto m_p m_e^3$ in the case of atomic matter, such as in the case of rocky planets, and $\rho\propto m_p m_\pi^3$ in the case of NS's. In virually all cases explored in the previous section, except for the case of galaxies, the typical $\rho$ is always determined by powers of fundamental mass scales, and essentially $\rho\propto m_{pl}^{4}$ if we define $m_{i}\equiv\lambda_i m_{pl}$ where $i$ stands for $e$, $p$, $\pi$, etc. The latter assumption, that 
particle masses can be presented as a fraction of the Planck mass indeed needs justification. 
This is provided in the context of a locally scale invariant formulation of GR in Section \ref{sec3.3}.
Since $\rho\propto m_{pl}^{4}\propto G^{-2}$, $G$ entirely vanishes from Equation 7.  

The vanishing of the dimensional $G$ from the equations in the primed system is intimately related to (and hints towards) scale-invariance of gravity. Notably, in the context of a fourth-order Weyl-invariant (WI) theory of gravity, the dimensional $G$ -- as any other dimensional constant -- is nowhere to be found in the governing action. Within the weak field approximation, the left-hand side of the Poisson equation is replaced by a double Laplacian, resulting in a $G$-independent gravitational potential with an external solution having a leading term $\propto r^{-1}$ in the spherically symmetric static case \cite{20}. This solution is $G$-independent, already in the primed coordinate system. Indeed, the relation of our proposed construction to another WI theory -- a locally scale invariant version of GR where $G$ is not absent but is rather promoted to a scalar field -- is discussed in Section \ref{sec3.3} below. 

Thus, in physical systems where density and potential are set by microphysics, and when analyzed in appropriately rescaled coordinates, the gravitational constant $G$ no longer shapes either the dynamics or the observable properties of the system.

\subsection{Generalization to Relativistic Systems}\label{sec3.2} 

The generalization to the relativistic case, applicable to, e.g., NS's, BH's, as well as the entire observable Universe, is straightforward.
Within the framework of general metric theories of gravity, given a metric field then gravitational redshift, gravitational lensing, etc. are derived from the geodesic equations with no recourse to $G$. The latter only appears at the Einstein field equations when we restrict ourselves to GR in presence of matter, relating curvature- to energy density-units.

Consider the Einstein-Hilbert (EH) action
\begin{eqnarray}
S_{EH}=\int \left(\frac{\mathcal{R}}{16\pi G}+\mathcal{L}_{m}\right)\sqrt{-g}d^{4}x, 
\end{eqnarray}
where $\mathcal{R}$ is the Ricci curvature scalar obtained from the metric 
$g_{\mu\nu}$ and its first and second derivatives in the usual way, 
$\mathcal{L}_{m}$ is the matter lagrangian density, $G$ is the universal gravitational constant and $g$ is the determinant of $g_{\mu\nu}$. 
In a wide class of matter configurations 
\begin{eqnarray}
\mathcal{L}_{m}\equiv G^{-\alpha}\mathcal{\bar{L}}_{m}, 
\end{eqnarray}
where $\mathcal{\bar{L}}_{m}$ is G-free, and $\alpha$ is a non-negative parameter. We theoretically motivate this particular form in Section \ref{sec3.3}, but for the purpose of the present section it is considered to be merely a convenient ansatz.   
Assuming that $g_{\mu\nu}$ is independent of $G$ (i.e. $\Phi$ is $G$-independent), and capitalizing 
on the fact that $\mathcal{R}$ contains terms of the form 
$g^{\alpha\beta}\partial^{2}g_{\mu\nu}$ and 
$\partial g^{\alpha\beta}\partial g_{\mu\nu}$, we then employ the coordinate transformation $x^{\mu}\rightarrow x'^{\mu}=x^{\mu}/\sqrt{G}$ to obtain
\begin{eqnarray}
S'_{EH}=\int \left(\frac{\mathcal{R}'}{16\pi}+G^{2-\alpha}\mathcal{\bar{L}}_{m}\right)\sqrt{-g}
d^{4}x', 
\end{eqnarray}
where $\mathcal{R}'$ is the Ricci curvature scalar computed in the usual way but now 
the derivatives are with respect to $x'^{\mu}$ rather than $x^{\mu}$. Consequently, under these assumptions the EH action can be brought into a manifestly G-free form in case that $\alpha=2$, 
where the physical significance of this particular case will be made clear below. 
In other words, the energy density, gravitational field, gravitational redshift, gravitational lensing, temperature of matter (e.g. plasma, photons, baryons, etc.), induced velocities, etc. and all other quantities derived directly or indirectly from $S'_{EH}$ are all manifestly 
G-independent. We will see below that G-independence implies that all these quantities are intensive rather than extensive, i.e. they are utterly independent of the complexity of the physical system, namely on the number of particles $N$ it is composed of. The reduced EH (rEH) action $\bar{S}_{rEH}\equiv\int \left(\frac{\mathcal{R}'}{16\pi}+\mathcal{\bar{L}}_{m}\right)\sqrt{-g}d^{4}x'$ (assuming that $\alpha=2$) then exposes the underlying microphysics that govern these self-gravitating systems. The Einstein equations can be then derived in the usual way, but this time with no $G$-dependence. In the case that $\mathcal{\bar{L}}_{m}$ describes perfect fluid then the temporal and spatial components of $T^{;\nu}_{\mu\nu}=0$ result in the continuity and Euler equations, respectively. Specifically, consider a perfect fluid with energy-momentum tensor 
$T_{\mu\nu} = (\rho + P) u_\mu u_\nu + P g_{\mu\nu}$,
where $\rho$ is the energy density, $P$ is the pressure, and $u^\mu$ is the 4-velocity.
Under the assumptions of small perturbations, i.e. 
$g_{\mu\nu} = \eta_{\mu\nu} + h_{\mu\nu}$ with $|h_{\mu\nu}| \ll 1$, weak gravitational field and non-relativistic velocities, perfect fluid with no viscosity or heat conduction, and linearization of Einstein equations, the Einstein field equations reduce to the Poisson equation for the gravitational potential, Equation 1, but with $G$ effectively set to unity, 
and the covariant conservation $\nabla_\mu T^{\mu\nu} = 0$ 
reduces to the classical fluid equations, Equations 2 and 3.
This result is general under the stated assumptions, and consequently $G$ entirely scales out from the classical fluid and Poisson equations.

\subsection{Relation to Local Scale Invariance}\label{sec3.3}

The $\mathcal{L}_{m}$ scaling with $G$ explored in Section \ref{sec3.2} naturally arises in the context of a WI version of the EH action. Specifically, it is a unique property of this WI action that the only $\mathcal{L}_{m}$, assuming it corresponds to a perfect fluid with EOS $w$, consistent with the underlying Weyl-invariance is $\mathcal{L}_{m}\propto|\phi|^{1-3w}$ where $|\phi|$ is the modulus of a complex scalar field, as we briefly discuss below. It is therefore proposed as a natural generalization of the discussion in Section \ref{sec3.2}. Specifically, the action is
\begin{eqnarray}
S_{WI}=\int[\xi|\phi|^{2}\mathcal{R}+\phi_{\mu}\phi^{*\mu}
+\mathcal{L}_{m}(|\phi|,\{\psi\})]\sqrt{-g}d^{4}x, 
\end{eqnarray}
where $\phi_{\mu}\equiv\frac{\partial\phi}{\partial x^{\mu}}$, 
$\mathcal{L}_{m}$ depends only on $|\phi|$ as well as other fields, 
collectively denoted as $\{\psi\}$, but not on the scalar field phase, and so the latter is a cyclic `coordinate' in field space, and $\xi=1/6$ guarantees that the free action is WI. The local non-conservation of energy-momentum is $T^{;\nu}_{\mu\nu}=\phi_{,\mu}\frac{\partial\mathcal{L}_{m}}{\partial\phi}+\phi^{*}_{,\mu}\frac{\partial\mathcal{L}_{m}}{\partial\phi^{*}}$. In addition, $T^{\mu\nu}$ satisfies a consistency relation 
$\phi\frac{\partial\mathcal{L}_{m}}{\partial\phi}+\phi^{*}\frac{\partial\mathcal{L}_{m}}{\partial\phi^{*}}=T$ where $T$ is the trace of the energy-momentum tensor. 
As shown in \cite{31}, the only 
$\mathcal{L}_{m}$ consistent with both constraints -- and assuming a multi-component perfect 
fluid where species $i$ is characterized by an equation of state $w_{i}$ -- is of the form 
$\mathcal{L}_{m}=\sum_{i}f_{i}(\{\psi\})|\phi|^{1-3w_{i}}$, where $f_{i}$ are $|\phi|$-independent. This form is analogous to the ansatz $\mathcal{L}_{m}=G^{2\alpha}\bar{\mathcal{L}}_{m}$ employed in Section \ref{sec3.2}. For the purposes of the present work is should be sufficient to fix the phase of 
$\phi$ to a constant value, $0$ by default.
Here, we naturally obtain this separation of the 
$|\phi|$-dependence from WI -- with no need to assume it -- hinting towards a profound relation between the microscopic/macroscopic duality on the one hand and local scale-invariance on the other hand. For convenience we define $\phi=\phi_{0}a(x)e^{i\chi}$ where the dimensionless function $a(x)$ is spacetime-dependent and $\chi$ is the phase. We further define the `rapidity' $\zeta$ via $a\equiv e^{\zeta}$ with $\zeta\in(-\infty,\infty)$, resulting in $\phi=\phi_{0}\exp(\zeta+i\chi)$. 
Renaming of the spacetime coordinates, $x^{\mu}\rightarrow |\phi|x^{\mu}$, allows recasting the 
action in the form
\begin{eqnarray}
S'=\int[\xi \mathcal{R}+\zeta_{\mu}\zeta^{\mu}+\chi_{\mu}\chi^{\mu}+\sum_{i}\lambda_{i}(\{\psi\})e^{-3(1+w_{i})\zeta}]\sqrt{-g}d^{4}x, 
\end{eqnarray}
where the action $S'$ is presented in the microscopic intensive frame, 
and $\lambda_{i}\equiv f_{i}(\{\psi\})\phi_{0}^{-3(1+w_{i})}$ are 
dimensionless. Note that we renamed the coordinates rather than carrying out coordinate transformation 
-- the latter would transform the metric field and the Ricci scalar in such a way that would 
leave the action invariant.
We note that the hallmark WI of the macroscopic frame is lost in the intensive frame. 
In addition, the action $S'$ is now formulated in an all-dimensionless scalar fields form as both 
$\zeta$, $\chi$ and the metric field $g_{\mu\nu}$ are purely dimensionless. In particular, the dimensional $G$ (or the related $\phi_{0}$) completely vanishes 
from the field equations derived from $S'$ if we limit ourselves to the case $w=-1$ where 
$\lambda$ reduces to a dimensionless number, independent of $\{\psi\}$.
Moreover, unlike Weyl transformations that apply to the fields, and the metric field in particular, that shift the compactness parameter, the coordinate transformation $x'^{\mu}\rightarrow|\phi| x^{\mu}$ employed in going from Equation 11 to 12 does not affect the fields, $g_{\mu\nu}$ in particular, i.e. the compactness is unchanged in going between the two coordinate systems, namely between the macroscopic and microscopic descriptions, or equivalently between the extensive and intensive descriptions. Two comments are in order here. First, we now see how the assumption made throughout the discussion below Equation 7 is justified; the case $w=-1$ corresponds to the 
form of mass densities of typical astrophysical systems considered in Section \ref{sec2} where 
$\rho\propto m^{4}$ and so masses that appear in $\rho$ or $\mathcal{L}_{m}$ 
can be defined in this context as $m_{i}=\lambda_{m}m_{pl}$ where $\lambda_{m}$ is a dimensionless parameter. This of course does not apply to inertial masses as we encounter neither necessity to introduce any changes to the standard model of particle physics, nor is it implied by the proposed construction. Rather, Equations 11 and 12 only imply that the scalar field $\phi$ governs both the Planck mass and the gravitational masses, not the inertial masses. The distinction in principle between inertial and gravitational masses is known and acknowledged but is effectively addressed 
by simply positing these two types of masses are the same. However, classically this is 
only a matter on convention. the second comment is of course the manifest invariance to the Planck mass in the case of $w=-1$. This bodes well with the fact that we model astrophysical systems in the present work by spherically symmetric, pure-gravity, static solutions. The only possible such configuration is the de-Sitter metric solution.

We now see how WI emerges; the dimensional $G$ 
in the macroscopic frame is merely the value of a dimensional scalar field -- `dressed' with WI then $G$ is promoted to a field ($\propto|\phi|^{-2}$). 
Once we rewrite the theory in the coordinate system $x'^{\mu}\equiv|\phi| x^{\mu}$, then WI is lost, and we essentially transformed from the macroscopic $N$-dependent description to a N-independent frame. Clearly, this logic could be reversed; starting from the non-WI single-particle dimensionless description, we transform to the multi-particle macroscopic dimensional description, where $|\phi|$ (i.e. the dimensional $G$) emerges, and so does WI manifests itself.  

It should be also stressed that the scaling of the various contributions to the energy density budget for all species changes by an overall $\propto|\phi|^{4}$ factor in going from the microscopic to the macroscopic description. For example, radiation rescales from $\propto|\phi|^{-4}$ to $|\phi|$-independent, dust rescales from $\propto|\phi|^{-3}$ to $\propto|\phi|$, 
and vacuum-like energy from $|\phi|$-independent to $|\phi|^{4}$. 
We then readily identify $a(x)$ in the microscopic system with the scale factor $a(t)$ 
of standard FRW expanding Universe model in the case of homogeneous and isotropic space. 
However, now we see why the latter is effectively G-independent and thus corresponds 
to a cosmological model described in an N-independent fashion.

Three cases are of special interest here. The first is $w_{i}=-1$ where the matter 
lagrangian in Equation 11 is $\zeta$-independent.
Although compact objects formally fit the $w = -1$ case in the static perfect fluid hard-sphere approximation, this is a simplification that does not reflect their true physical nature. This approximation arises because these objects are tightly packed, with number densities directly determined by fundamental masses -- such as one particle per Bohr volume for rocky planets or one particle per nucleon volume in NS's. Consequently, their typical energy densities scale as mass to the fourth power, $\rho \propto m^{4}$, corresponding to $w = -1$. Physically, a pure-gravity static spherically symmetric hard-sphere matter configuration corresponds to $w=-1$ because it is balanced against its own gravity by the negative pressure. In a more realistic model other interactions are involved as well, and in addition the density and pressure profiles are r-dependent, 
consistent with $w\approx 0$.
In contrast to compact objects, cosmological baryon number density $n_{b}$ is not fixed by any mass scale but is {\it a priori} a free parameter that is observationally determined, so the energy density scales linearly with mass in this case, 
$\rho \propto m$, corresponding to $w = 0$. Thus, the $w = -1$ classification for compact objects is a formal artifact of the simplified model, reflecting the fundamentally different physics governing their tightly packed structure compared to the diffuse cosmological medium.

The second case corresponds to galaxies. In this case $\rho\propto G\propto |\phi|^{-2}$, which corresponds to $w=1$, i.e. an EOS of stiff incompressible matter. This is again an artifact of the 
approximations and assumptions made in deriving the typical galactic energy density -- a more realistic description with $w\approx 0$ requires the inclusion of non-gravitational pressure 
sources and departure from the hard sphere approximation employed in \cite{15,16}.

Third, in the homogeneous and isotropic case, i.e. $a=a(t)$, the FRW equation on 
a static background where the entire evolution is that of the Planck mass trivially follows from 
variation of the action in Equation 12 with respect to $\zeta$ under the assumption that $\chi$ is fixed to 0. The details of such variation closely follow that of \cite{31}.
The interpretation of the observed universal expansion is as follows. As we peer to ever larger distances $\phi$ is correspondingly smaller, relatively increasing the $w=0$ and ultimately the $w=1/3$ components, analogous to the conventional expanding Universe interpretation. Much like in the SM of cosmology, $a(t)\propto (t/t_{0})^{1/2}$ or $(t/t_{0})^{2/3}$ in the radiation- and matter-dominated cosmological eras. Since the present time, $t_{0}$, is arbitrary, and is clearly $G$-independent, then cosmological redshift is $G$-independent, much like as in the case of compact objects discussed in Section \ref{sec2}. In fact, the Friedmann equation is written in a G-independent fashion $H(z)=H_{0}\sqrt{\Omega_{r}(1+z)^4+\Omega_{m}(1+z)^3+\Omega_{\Lambda}}$ where $H_{0}$ 
is the present-day expansion rate, and 
$\Omega_{r}$, $\Omega_{m}$ and $\Omega_{\Lambda}$ are the radiation, matter and DE density parameters of radiation, matter and DE, respectively, given in critical density units. 
This equation is normally fitted to observational data with no reference to $G$. Only when 
combined with direct measurements of baryon abundance in the local Universe, or with independently measured CMB temperature, coupled with the (observationally strongly motivated) assumption that the CMB is black body radiation is $\Omega_{r}$ independently estimated resulting in a weak or moderate handle on $G$, e.g. \cite{33}.

\section{Implications of Extensive Planck Mass}\label{sec4}

Having experimented with the intensive nature of the compactness parameters of typical astrophysical systems, $\Phi$, and with the emergence of WI version of GR and the emergence of a 
macroscopic description starting from a microscopic $G$-independent description, we are 
conveniently positioned to explore the implications of the possibility that the experimentally inferred Planck mass is actually an extensive property of astrophysical systems.

In purely Cavendish‐type tests only the combination $Gm_{1}m_{2}$ is being measured. Notably, under arbitrary rescaling $G\rightarrow Gk^{2}$ and $m_{i}\rightarrow m_{i}k^{-1}$ where $k$ is an arbitrary, universal rescaling function, the measured $Gm_{1}m_{2}$ remains invariant. The conventional choice, $k=1$, corresponds to the imposition that 
gravitational mass that appears in the universal law of gravitation is identified with the inertial mass. It is then possible to nominally define the Planck mass by 
$m_P^{(\rm nom)}=\sqrt{\frac{\hbar\,c}{G_{\rm k=1}}}\sim10^{19}\,\mathrm{GeV}$.
However, this nominal value is purely convention‐dependent. 
Notably, this `scale invariance' of $Gm_{1}m_{2}$ is actually naturally built in the WI version of GR that is discussed in Section \ref{sec3.3} where the modulus of the scalar 
field $\phi$ regulates both the Planck mass and gravitational masses in a way that leaves 
the coupling strength $Gm_{1}m_{2}=m_{1}m_{2}/m_{pl}^{2}$ invariant.

As discussed above, a subtle but significant aspect of gravitational physics is the fashion in which the gravitational constant $G$ enters observable phenomena. In many gravitational systems -- particularly those that are virialized, such as stars, planets, and galaxies -- upper bounds on quantities accessible to observation (e.g., gravitational redshift, line broadening, lensing deflection angles, escape velocities, and virial velocities) are determined by microphysics with no reference to $G$. 
It is only in the case of non-virialized systems, or subsystems of otherwise virialized (or degeneracy pressure-supported systems) systems, or in the direct interaction between two effectively pointlike masses (e.g., the Earth–Sun system), that $G$ appears explicitly in observable quantities, as in Newton gravitational law. In these cases, the force or acceleration is directly proportional to $G$ and can be measured unambiguously.

This pattern suggests that the role of $G$ is, in a sense, emergent; it is hidden in configurations where gravitation is balanced by other non-gravitational interactions and revealed only when comparing/measuring gravitational effects to/by non-gravitational standards or when considering pointlike interactions. This conclusion is ingrained and manifested throughout the analysis in Section \ref{sec3}.
In this context, $G$ serves as a conversion factor between energy density that is set by microphysics and the macroscopic spacetime curvature, but its absolute value is often inaccessible in the absence of a fixed external non-gravitational scale. This observation highlights a profound feature of gravitational physics, with possible implications for our understanding of the fundamental vs. emergent nature of gravity.

It should be stressed that the absolute value of $G$ is primarily determined by laboratory experiments on Earth, such as Cavendish-type measurements, and even then it is the measurement of gravitational interaction by comparison to non-gravitational forces that allows the empirical determination of $G$. Solar system observations -- including lunar laser ranging, planetary orbit tracking, and spacecraft Doppler data -- measure the product $GM$ (where $M$ is the mass of a celestial body) and cannot independently determine $G$ without knowing $M$. Therefore, these methods mainly constrain temporal or spatial variations of $G$ relative to the Earth-based value, rather than empirically inferring its absolute magnitude.

The following example illustrates one way in which $G$ might emerge.  As discussed 
below Equation 6 -- and shown with specific cases in Table \ref{table1} -- systems whose characteristic density $\rho$ is fixed by microphysics (and does not involve $G$) exhibit scalings $R \propto G^{-\frac{1}{2}}$ and $M \propto G^{-\frac{3}{2}}$. By contrast, systems (like galaxies) whose formation process depends on $G$ have $\rho\propto G$, leading instead to $R \propto G^{-1}$ and $M \propto G^{-2}$ as required by the $G$-independence of $\Phi$.

Here we focus on the more generic case where $\rho$ truly does not depend on $G$, and in particular on habitable planets. Our objective is to explain -- at the dimensional analysis level -- 
why terrestrial Cavendish-type experiments measure an effective Planck mass 
$m_{pl}\sim 10^{19}$ GeV rather than some other value. From Table \ref{table1}, one finds that the typical radius of such a planet is $R\sim\frac{\sqrt{\epsilon\alpha_{e}}m_{pl}}{m_{p}}a_{0}$ where $a_{0}$ is the Bohr radius, $\alpha_{e}$ the fine-structure constant, and $\epsilon$ a parameter of order $O(0.1)$ as was already defined in Section \ref{sec2} but we reiterate for convenience.  Since atoms in a rocky planet are tightly packed, we also have $R\sim(2a_{0})N^{1/3}$ where $N$ is the number of atoms. Combining these two estimates to eliminate $R$ we arrive 
at $m_{pl}=\frac{2m_{p}}{\sqrt{\epsilon\alpha_{e}}}N^{1/3}$.

Conventionally, we would choose $k=1$ as defined at the beginning of this section, i.e. we set the gravitational mass equal to the inertial mass thereby fixing $G$, and interpret this relation as saying that the number of atoms $N$ in a habitable planet is determined by $m_{pl}$, $m_{p}$, 
$\alpha_{e}$ and $\epsilon$. Indeed, a similar approach has been 
adopted in the context of estimating the number of protons in a typical nuclear burning star, e.g. \cite{21,11,22}.  
However, adopting the viewpoint that $m_{pl}$ is merely an emergent conversion factor, and in addition noting that empirically
$\tfrac{2\,m_{p}}{\sqrt{\epsilon\,\alpha_{e}}}$ is of order the Higgs mass $m_{H}$, this logic can be reversed, and we are led to tentatively conjecture instead that
\begin{eqnarray}
m_{\rm pl}\sim m_{H}\,N^{1/3},  
\end{eqnarray}
i.e. that $m_{\rm pl}$ is determined by $N$ and not the other way around -- the larger 
$N$ is the weaker is gravity, and the larger is $m_{pl}$.
From this perspective, the Earth-based laboratory determination of a 
small $G$ (equivalently a large $m_{pl}$) simply reflects the fact that 
typical habitable planets like Earth contain 
$N\sim \epsilon^{\frac{3}{2}}\lambda^{\frac{3}{2}}\sim 10^{51}$ atoms, producing a $\sim17$-order-of-magnitude hierarchy $m_{\rm Pl}/m_{H}\sim N^{1/3}$. Following this logic and repeating Cavendish-like experiments on, e.g. the moon or Mars (while fixing $k=1$) 
will result in different inferred values for $G$ (i.e. values different from that inferred 
on Earth) if our assumption that $G$ is an 
emergent value following Equation 13 rather than a universal constant is correct. 
This is a falsifiable prediction of our proposal. It should be also emphasized that 
the proposed Planck-to-Higgs mass relation, Equation 13, is not rigorously derived 
from any first princples and it may well be an empirical coincidence that it 
empirically fits Earth-based measurements. However, the 
scaling $m_{pl}\propto N^{1/3}$ is more robust as was just discussed above Equation 13, 
and the proprtionality factor can be some other non-Higgs scale. 
Nevertheless, it would be a remarkable coincidence 
if the latter just happens to nearly coincide with $m_{H}$.

Even the number of particles in Earth is determined 
by fundamental principles to be $N\sim \epsilon\lambda^{\frac{3}{2}}$. Fitting the order-of-magnitude Earth compactness $|\Phi|\sim \epsilon\alpha_{e}^2\mu^{-1}$ (as shown in Table \ref{table1}) to its measured compactness we obtain that $\epsilon \sim 0.05$. Combining the proposed scaling $m_{pl} \propto N^{1/3} m_H$ with $N\sim \epsilon\lambda^{\frac{3}{2}}$ we than conclude that $m_{p}\sim\sqrt{\epsilon\alpha_e}m_{H}$. Assuming $\epsilon \sim 0.05$ and the measured proton mass we remarkably obtain the order-of-magnitude estimate $m_{H}\sim 50$ GeV.
Crucially, within each system, the ratio $m / m_{pl}$ is independent of $N$ because both gravitational mass and Planck mass scale identically with $N^{1/3}$. This ensures gravitational interactions between any two systems with different particle numbers remain universal and consistent with the equivalence principle, as the gravitational coupling between two systems with masses $m_{1}$ and $m_{2}$ depends only on these ratios as 
$\alpha_{g}^{1,2}=Gm_{1}m_{2}=\frac{m_{1}}{m_{pl}}\frac{m_{2}}{m_{pl}}$. 

Thinking in terms of current terrestrial experimental standards and precision of $G$-inference 
could be disheartening. Indeed, conducting a Cavendish-like experiment on, e.g. the moon or Mars is a daunting task that presents both intriguing scientific opportunities and significant challenges. 
Although technically demanding, successfully conducting this experiment could provide significant insights into the nature of gravity, making it a scientifically valuable endeavor worth pursuing despite the obstacles. This is especially true given that falsifying either the universality of $m_{pl}$ or its proposed $\propto N^{\frac{1}{3}}$ dependence on, e.g., the moon is not as demanding as a full-fledged precision inference of $G$ with current standards of precision measurements. Specifically, according to Equation 13 the fractional relative divergence of $m_{pl}$ inferred in two systems that contain $N_{1}$ and $N_{2}$ particles respectively is expected to be $\Delta m_{pl}/m_{pl}=1-(N_{2}/N_{1})^{\frac{1}{3}}$. With the moon-to-Earth mass ratio of $\sim 0.012$ and assuming small binding energies in both systems we obtain 
$\Delta m_{pl}/m_{pl}\sim 0.77$, and consequently a very modest $\sim 10$ percent precision is already more than sufficient to falsify the proposed scaling.

Even on Earth $G$ can vary from place to place on its surface, because the surface of planets is shaped not only by gravity but also by the electromagnetic interaction as well, i.e. by the strength of matter as measured by its bulk modulus (compressibility). 
On planets, this is typically of order $Ry/(2a_{0})^{3}$, e.g.  \cite{14}. The estimated mass of Mount Everest, approximated  as a cone with a base radius of about 5 km and height 8.8 km, and assuming an average Earth density of $5 gr/cm^{3}$, is $\sim 10^{18} gr$. Comparing this to Earth mass $5.97\times 10^{27} gr$ and assuming similar mass densities for both we find 
$\Delta N/N=O(10^{-10})$, the fractional Planck mass difference between sea level and the summit of mount Everest is then $\Delta m_{pl}/m_{pl}\approx\Delta N/(3N)$ which is smaller than one part in ten billion. This is much smaller a difference than current precision in $G$ measurements. The latter is not much better than one part in $10^{5}$. Thus, whereas a moon-based Cavendish-like experiment can put Equation 13 to test, near-future Earth-based falsification of this conjecture seem to be not feasible.

One immediate consequence of Equation 13 is that the Higgs-to-Planck energy-density ratio becomes
\begin{eqnarray}
\rho_{vac}\sim m_{pl}^{4}\propto N^{4/3}.
\end{eqnarray}
On cosmological scales where $N=O(10^{88})$ is dominated by CMB photons and primordial neutrinos
this ratio is $O(10^{120})$. This is tentatively close to the $\sim$122 
orders-of-magnitude fine tuning of the observed $\rho_{\Lambda}$, i.e. 
the energy density associated with the cosmological constant, and the naive expectation 
for the zero-point vacuum energy density, $\rho_{vac}=O(m_{pl}^{4})$ \cite{27}. From the present work perspective, and as we explicitly saw in 
Section \ref{sec3}, the cosmological description of a cosmic evolution dominated by $w=-1$, 
i.e. a vacuum energy-like cosmological constant, is in the microscopic, intensive, frame that corresponds to $N=1$. Compared to the extensive frame description with $N=O(10^{88})$, 
the $\sim 122$ orders of magnitude gap between observations and naive theoretical 
expectations can be almost fully bridged by realizing that the apparent mismatch results from 
comparing intensive to extensive quantities in a very complex system that contains 
$O(10^{88})$ particles.

Although the extensive vs. intensive perspective requires a markedly unusual treatment of $G$, 
that admittedly diverges even from other emergent-gravity perspectives, it offers a new 
perspective -- by virtue of its $N$-dependence -- on both the Planck-Higgs mass hierarchy 
and the `unexpectedly' small cosmological constant.

At first glance, it may seem that we have simply traded the Higgs-Planck mass hierarchy and the 
cosmological constant problems by the question; why is our Universe so vast in complexity, 
i.e. why do habitable planets, nuclear-burning stars, and the observable Universe contain $N\sim10^{51}$, $10^{57}$ and 
$10^{88}$ particles, respectively? However, our interpretation offers a unified, quantitative scaling of $G$ with system complexity $N$, thereby bringing both hierarchies under one umbrella.  The ultimate (qualitative, if not quantitative) resolution of ``why so large complexity?'' 
may ultimately -- perhaps unsurprisingly -- be anthropic; only a sufficiently complex cosmos can give rise to complex observers.

\section{Summary}\label{sec5}

Typical astrophysical velocities are almost always non-relativistic, and the gravitational 
wells $\Phi$ characterizing these astrophysical systems are correspondingly shallow, 
i.e., $|\Phi| = r_{s}/R \ll 1$, where $r_{s}$ and $R$ are their typical Schwarzschild radius and actual size, respectively. According to the generalized Second Law of thermodynamics, this implies that the entropy budget of our Universe is much smaller than the holographic entropy bound, $S\sim 10^{122}$ \cite{32}.

This can be neatly explained by the fact that astrophysical systems are either regulated by cooling processes (governed by the electromagnetic interaction) or are required not to allow certain electromagnetic processes to occur. In other words, these systems must remain below specific temperature thresholds, otherwise their virial velocities would exceed their escape velocities and the systems would rapidly disperse. Typical temperatures are set by atomic or molecular physics, with $T \sim m_{e} \alpha_{e}^{2}$, so that virialized objects are typically characterized 
by $|\Phi| \sim \alpha_{e}^{2} \mu^{-\gamma}$, where $\mu = m_{p}/m_{e}$ and $\gamma$ is positive.

Had $\alpha_{e}$ and $\mu$ been of order unity, the entire Universe would be teeming with BH's and NS's, with typical velocities approaching the speed of light, shortly after the Big Bang -- regardless of its initial entropy state. So why is the Universe not predominantly populated by BH's and NS's? Simply because $\alpha_{e} \ll 1$ and $\mu \gg 1$.

Whereas the reasoning laid out in this work is straightforward, the somewhat surprising conclusion -- that although the existence of astrophysical systems depends on $G$ and specifically on its empirical value, large-scale velocities at the system boundary are determined solely by microphysics -- essentially by the electromagnetic interaction rather than by gravitation -- is not sufficiently appreciated in the literature. One objective of this work was to emphasize this rather surprising fact. Although order-of-magnitude estimates of $M$, $R$ and $T$ for various astrophysical systems from first principles are discussed in the literature, the facts 
that $|\Phi|$ is small due to the weakness of the electromagnetic interaction and the small electron-to-proton mass ratio, and $|\Phi|$ is independent of $G$, seem to be overlooked, or at least not sufficiently acknowledged. In particular, we stress that this $G$-independence 
of $|\Phi|$ applies specifically to entire, virialized, naturally-formed astrophysical systems -- such as stars, planets, and galaxies -- that have formed and equilibrated through self-gravity and the relevant microphysical processes. Arbitrary non-virialized subsystems, like a piece cut from a planet, do not exhibit this property.

In gravitational systems, distinguishing between intensive and extensive quantities offers a coherent framework to tentatively address the mass hierarchy and dark energy problems. Specifically, the inertial mass scale $m_H$, set by particle physics, is a constant of nature, 
independent of system size or particle number $N$. In contrast, we put forward the proposal 
that the gravitational mass scale, characterized by the effective Planck mass $m_{pl}$, 
scales as $m_{pl} \sim m_H N^{1/3} $ (ignoring possible geometric factors -- or others -- of order unity), a conjecture that could be directly falsified by future moderate-precision moon-based Cavendish-like 
experiments. For example, Earth’s particle content $N \sim 10^{51}$ quantitatively explains the observed hierarchy $m_{pl} / m_H \sim 10^{17}$ measured in Cavendish experiments, linking microscopic particle masses to macroscopic gravitational scales. 
Extending this to cosmology, vacuum energy density $\rho_{vac}$ is measured as we saw in Section \ref{sec3} as an intensive quantity, i.e. $\rho_{vac} \propto m_{pl}^4 \propto N^{4/3}$. With $N \sim 10^{88}$, approximately the number of CMB photons in the observable Universe, this scaling naturally suppresses naive particle physics predictions by about 120 orders of magnitude; observing the Universe from within, cosmological measurements probe intensive quantities that reflect the gravitational scaling tied to the Universe particle content, e.g. the energy density rather than total mass of the Universe (much like the energy density of the CMB does not depend on the total number of CMB photons in the Universe and the mass densities of e.g., planets, WD and NS depend exclusively on microphysics). As a consequence, the `naturalness' problems associated with the mass hierarchy and cosmological constant could be rephrased as why is our Universe so complex, i.e. why, e.g. planet Earth and the observable Universe contain $O(10^{51})$ and $O(10^{88})$ 
particles, respectively. This question is rarely asked in the context of these puzzling problems.

\section*{Acknowledgments}
Anonymous referees are acknowledged for their constructive comments that 
stimulated a significant revision and expansion of the scope of this work.
This research was supported by a grant
from the Joan and Irwin Jacobs donor-advised fund at the
JCF (San Diego, CA).

\end{document}